\documentclass[preprint]{aastex}
\usepackage{emulateapj5}
\begin{document}
\title{Carbon Monoxide in the Type~Ic Supernova 2000ew \protect\footnotemark}

\shorttitle{CO in the Type~Ic SN~2000ew}
\shortauthors{Gerardy et al.}

\author{Christopher L. Gerardy \& Robert A. Fesen}
\affil{6127 Wilder Laboratory, Physics \& Astronomy Department \\
       Dartmouth College, Hanover, NH 03755, USA}
\author{Ken'ichi Nomoto \& Keiichi Maeda}
\affil{Department of Astronomy and Research Center for the Early Universe\\
University of Tokyo, Bunkyo-ku, Tokyo 113-0033, Japan}

\and
\author{Peter H\"oflich \& J. Craig Wheeler}
\affil{Department of Astronomy, University of Texas at Austin, Austin, TX 78712}

\footnotetext{Based in part on data collected at Subaru Telescope, which is 
operated by the National Astronomy Observatory of Japan.}

\begin{abstract} We present \textit{K}-band (1.9 -- 2.5 \micron) spectra of the Type~Ic SN~2000ew observed with IRCS on the Subaru Telescope.  These data show the
first detection of carbon monoxide (CO) emission in a Type~Ic
supernova.  The detection of CO in SN~2000ew provides further evidence that
molecule formation may be a common occurrence in core-collapse supernova ejecta.
The spectrum also contains narrow emission lines of [\ion{Fe}{2}] and 
\ion{He}{1}\ probably from dense clumps of hydrogen-poor circumstellar gas 
surrounding SN~2000ew.  Our spectrum of SN~2000ew shows no trace of an 
unidentified feature seen near 2.26~\micron, just blueward of the CO emission, 
in the spectrum of SN~1987A and we discuss proposed detections of this feature 
in other Type~II supernovae.
\end{abstract}


\section{Introduction}
The detection of first-overtone carbon monoxide (CO) emission near 2.3~\micron\
in SN~1987A opened exciting new possibilities for the study of supernovae (SNe).
Unfortunately, for nearly a decade, SN~1987A remained the sole supernova with 
detected molecular emission. However, with the maturing of near-infrared (NIR) 
spectrographs, observations of several SNe in the late-time nebular phase have
been made.  As a result, CO emission has now been detected in other Type~II 
supernovae: SN~1995ad \citep{spyromilio96}, SN~1998S \citep{gerardy00,fassia01},
SN~1998dl, and SN~1999em \citep{spyromilio01}.  These observations suggest that 
molecule formation may be a common occurrence in Type~II SNe.  

The study of molecular emission in SNe can provide valuable information about 
the composition, explosion dynamics, and the late-time temperature evolution and
energy balance in the ejecta.  For example, the detection of CO formation can 
place constraints on the mixing in the SN ejecta.  Because CO is quickly 
destroyed by the presence of ionized helium, the detection of CO emission in a 
Type~II or Ib supernova implies either that the CO is not microscopically mixed 
with helium or that the helium is not ionized \citep{lepp90,gearhart99}.  In the
case of a Type~Ic supernova there is likely little or no helium left in the 
outer envelope at the time of core collapse.  As a result, CO emission in a 
Type~Ic does not place strong constraints on the mixing between ejecta layers.

Due to its very large number of collisionally excitable energy levels, CO 
can be an important coolant of the SN ejecta.  Indeed, at temperatures of a few 
thousand K, CO emission may be the dominant cooling mechanism 
\citep{HSZ89,liu92,spyromilio96,liu95,gearhart99}. 

CO cooling may, in turn, play a key role in dust formation in supernova 
ejecta.  SN~1987A and SN~1998S, two supernovae with detected CO emission, also 
showed strong evidence of dust formation in the ejecta.  In SN~1987A, thermal 
emission from dust was detected in the mid and far-infrared while, at the same 
time, line emission from the ejecta shifted to the blue (\cite{MC93}, and 
references therein).  Similarly, late-time spectra of SN~1998S exhibited
H and He lines with a multi-peak line profile, the red side of which faded
dramatically with time, while the blue side remained nearly constant.
In both cases, the changes in line emission were attributed to the formation
of dust in the ejecta, which obscured the far side of the supernova causing the
redshifted emission to fade.  

Thermal emission from dust was also seen in SN~1998S \citep{gerardy00,fassia01},
but this was not due to dust formation in the ejecta.  In this case, the 
emission was likely from pre-existing dust in an extended cloud which was heated
by X-rays and UV light from a strong interaction between the supernova and dense
circumstellar gas \citep{gerardy02}.

Analysis of near-infrared CO emission can also provide information about the
CO-rich ejecta.  For example, the overall shape of the CO emission profile can 
be used as a temperature diagnostic for the CO emitting gas
(e.g.~\cite{spyromilio88,
SH89,liu92}).  The fine structure in the near-infrared is formed by 
transitions between different vibrational levels, with higher level transitions 
emitting at longer wavelengths.  As the temperature decreases, the higher 
vibrational levels become de-populated and emission from the red end of the CO 
profile decreases. 

For high S/N data, it is possible to determine the velocity of the CO emitting 
gas from the shape of the CO profile \citep{gerardy00,fassia01}.  At low 
velocities, the band structure of the CO emission is well defined and the blue 
edge of the emission profile is quite sharp.  As the expansion velocity 
increases the bands blend together, smearing out the pattern of peaks and 
troughs.  Also, the short-wavelength end of the profile creeps farther toward 
the blue and the rise becomes progressively shallower.  

High velocity CO can have important implications for the progenitor star.  To 
accelerate a significant amount of CO out to high velocity in a Type~II 
supernova, the progenitor must be massive ($M \geq 20 M_\odot$ for $V_{\rm CO} 
\geq 3000$ km~s$^{-1}$) and must have lost a large portion of its H and He rich 
mantle prior to core-collapse \citep{gerardy00}.  The large progenitor mass
is required to build up a large carbon/oxygen layer which will lie far enough
out in the ejecta to accelerate to high velocity.  In a smaller progenitor,
the C/O rich layer will be buried too deeply beneath the outer mantle.  

On the other hand, for stripped-envelope supernovae, where the pre-collapse mass
loss becomes extreme, the requirement for a massive progenitor may be relaxed.
For instance, in a Type~Ic supernova there is likely little or no mass left 
above the C/O rich layer and quite large C/O velocities might be expected from a
fairly low mass progenitor. However, if the C/O velocity becomes too high, then 
the density might drop too fast for a significant amount of CO to form. 

Here we present the first detection of CO in a Type~Ic supernova, SN~2000ew.
In \S~2, we describe the observations and data reduction.  In \S~3 we 
present
the data and compare the CO emission observed in SN~2000ew to that seen in the 
NIR spectrum of SN~1987A.  We also discuss the conclusion of 
\citet{spyromilio01} that both CO emission and an unidentified feature seen in 
the spectrum of SN~1987A near 2.26~\micron\ are ubiquitous features in the
NIR spectra of Type~II SNe. 

\section{Observations}
SN~2000ew was discovered in NGC~3810 on 28 Nov 2000 by T. Puckett and A.  
Langoussis with the Puckett Observatory 0.30-m automated supernova patrol
telescope \citep{puckett00}.  SN~2000ew was originally classified as a Type~Ia 
supernova by \citet{dennefeld00}, but was subsequently found to be of Type~Ic
\citep{filippenko00}.  

According to VSNET\footnote{http://www.kusastro.kyoto-u.ac.jp/vsnet/} white 
light magnitude estimates, it reached a maximum of $\approx 14$ mag around 4 
Dec 2000.  The recession velocity of NGC~3810, corrected for Virgocentric 
infall, is 1053 km~s$^{-1}$ (taken from the Lyon-Meuden Extragalactic 
Database\footnote{http://leda.univ-lyon1.fr/}.)  Using H$_0 = 72$ 
km~s$^{-1}$~Mpc$^{-1}$ \citep{freedman01} yields an approximate distance to 
SN~2000ew of 15 Mpc and implies a maximum light brightness of $\approx -17$ 
mag.  This is somewhat faint for Type~Ic supernovae which typically show peak
magnitudes around $M_V \approx -17.5$ -- $-18$ \citep{clocchiatti00}.  However 
with the crudeness of our magnitude estimate it is difficult to determine if 
SN~2000ew is unusually sub-luminous or just on the faint end of the luminosity 
spectrum for Type~Ic SNe.

On 5 Mar 2001, we obtained a 1.9--2.5 \micron\ spectrum of SN~2000ew using
the Infrared Camera and Spectrograph (IRCS; \cite{kobayashi00}) on the 8.2~m
Subaru Telescope.  A 0\farcs6 slit was used yielding a resolution of 
about 850 km~s$^{-1}$. The spectrum was collected in eight 400~s exposures
dithered between two positions along the slit in an ABBA pattern.  First order 
sky subtraction was obtained by subtracting `B' frames from `A' frames.  1-D 
spectra were then extracted using standard IRAF tasks.   The spectra were 
wavelength calibrated using an argon lamp spectrum obtained at the end of the 
night, and the calibration was tested by measuring the wavelengths of night-sky 
lines in the data. 

The data was corrected for telluric absorption using nearby AV and GV
stars selected from the Gemini Spectroscopic Standard Star 
Catalogues\footnote{www.us-gemini.noao.edu/sciops/instruments/niri/NIRIIndex.html}.  
The G-dwarf was divided by a solar spectrum 
\citep{solar1,solar2}\footnote{NSO Kitt Peak FTS data used here were
produced by NSF/NOAO} to remove stellar features.  The result was used to 
correct for telluric absorption in the A dwarf spectrum.  The stellar features 
were fit from the corrected A-star spectrum and removed from the raw A-star 
spectrum.  The raw A-star spectrum with the stellar features removed was then 
used to correct the target data for telluric absorption.  (See 
\cite{HCR96,hanson98} for a more detailed discussion of this technique.) The 
instrumental response was removed by matching the observed spectra of the A-star
telluric standards to model spectra from the stellar atmosphere calculations of 
\citet{kurucz94}.

\begin{figure*}[t]
\begin{center}
\includegraphics[scale=0.50,clip=true,trim=0bp 20bp 0bp 0bp,angle=270]{%
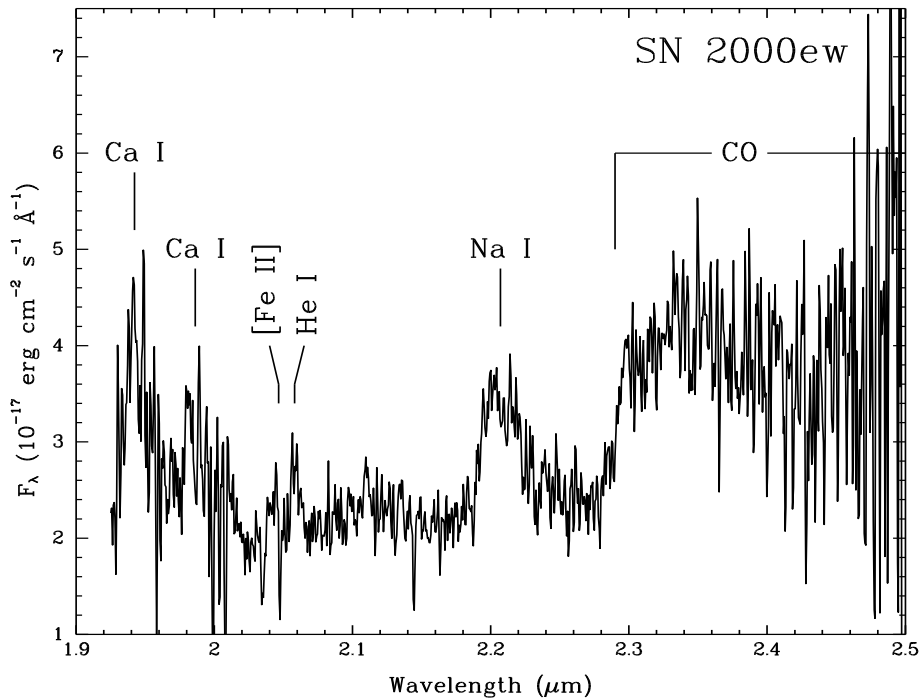}
\end{center}
\caption[Subaru \textit{K}-band spectrum of SN~2000ew taken on 5 Mar 2001]{The
1.9 -- 2.5~\micron\ spectrum of SN~2000ew taken on 5 Mar
2001 with IRCS on the 8.3~m Subaru telescope.  The spectrum shows broad
emission from \ion{Ca}{1}, \ion{Na}{1} and first-overtone
carbon monoxide emission as well as narrow emission from [\ion{Fe}{2}] and
\ion{He}{1}.  The wavelengths shown are in the rest frame of the host galaxy
NGC~3810.\label{sn00ew_fig1}}
\end{figure*}

The absolute flux level was set using NIR photometry taken a week later on
12 Mar 2001 using TIFKAM \citep{pogge98} on the 2.4~m Hiltner telescope at MDM 
Observatory.  This night was not photometric and the SN brightness was measured 
relative to a field star 93\arcsec~S and 26\arcsec~E of the 
supernova.  Boot-strapping the NIR photometry of this star from calibrated 
images taken in Jan 2001 resulted in $K = 15.34 \pm 0.10$ for SN~2000ew on 12 
Mar 2001.  This was then converted to flux units using the zero-points of 
\citep{schultz01}, and the flux was set by matching the flux of the spectrum 
averaged over the combined filter and atmospheric bandpass of 
\citep{hawarden01}.  We estimate that the absolute fluxing of the spectrum is 
accurate to 20\%.

\section{Discussion}
Figure~\ref{sn00ew_fig1} shows the full 1.9--2.5~\micron\ spectrum of SN~2000ew 
on 5 Mar 2001, 
some 97 days after discovery and about 90 days after maximum light.  The 
spectrum has been Doppler shifted to remove the 993 km~s$^{-1}$ heliocentric
radial velocity of NGC~3810 \citep{haynes98}.

Strong emission from the first-overtone band of CO is seen from 2.29~\micron\ to
the red end of the spectrum at 2.5~\micron. The spectrum also shows three broad
emission features near 1.94, 1.98 and 2.22~\micron.  We identify the 1.94 and 
1.98~\micron\ features as emission from the \ion{Ca}{1} 
4p$^3$P$^{\rm o}$--3d$^3$D 
multiplet, and the 2.22~\micron\ feature as \ion{Na}{1} 
4s$^2$S--4p$^2$P$^{\rm o}$.  
These features were identified in the spectrum of SN~1987A \citep{meikle89}.  
All three lines have a FWZI of about 4000 km~s$^{-1}$, peak around --500 
km~s$^{-1}$ and exhibit relatively boxy profiles that fade slightly from the 
blue to the red. 

In addition to these broad emission lines, the spectrum also contains two 
narrow emission features at 2.04 and 2.06 \micron\ which we identify as 
[\ion{Fe}{2}] 
2.0466~\micron\ and \ion{He}{1} 2.0587~\micron, respectively.  Both lines are 
unresolved and the narrowness of these features suggests that they are 
associated with circumstellar emission rather than supernova ejecta.  Both lines
have been identified in clumps of dense circumstellar gas in the Cas~A supernova
remnant, along with two other lines in the spectral region covered here: 
\ion{H}{1} 2.1661~\micron\ (Br$\gamma$) and [\ion{Fe}{2}] 2.2244~\micron\ 
\citep{gerardy01}.  The
latter [\ion{Fe}{2}] line would be lost in the broad \ion{Na}{1} feature, but 
the Br$\gamma$ 
line lies in a relatively empty region of the spectrum and no trace of this line
can be seen in our spectrum of SN~2000ew.  This suggests that the circumstellar 
clumps detected here may be hydrogen poor, perhaps originating in the He mantle
of the progenitor star and shed prior to core-collapse.  Such an interpretation 
is consistent with the general picture that Type~Ic progenitors shed a 
substantial portion of their helium rich mantle prior to core collapse 
(\cite{nomoto95}; \cite{filippenko97}, and references therein).  
The lack of hydrogen emission also suggests that the narrow lines are not 
associated with a background \ion{H}{2} region or other ISM gas in the host 
galaxy.
 
\subsection{Significance of CO emission}

\begin{table*}[t!]
\caption{CO detections in NIR spectra of core-collapse SNe}\label{sn00ew_tab1}
\begin{center}
\begin{tabular}{lcccc}
\hline \hline
Event & SN Type & Epoch of last &  Epoch of first & Reference\\
      &         & CO non-detection & CO detection \\
\hline
\multicolumn{5}{c}{\bf CO Detections}\\
\hline
SN~1987A & IIP Pec & 112~d   & 192~d & 1 \\
SN~1995ad & IIP    &         & 105~d & 2 \\
SN~1998S & IIn     &  ~44~d    & 109~d & 3 \\
SN~1999dl & IIP    &         & 152~d & 4 \\
SN~1999em & IIP    & 118~d   & 178~d & 4,5 \\
SN~1999gi & IIP    & ~74~d    & 126~d & 5 \\
SN~2000ew & Ic     &  ~39~d    & ~97~d  & 6,7 \\
\hline
\multicolumn{5}{c}{\bf Non-Detections}\\
\hline
SN~1990W & Ic      &  ~90~d    &       & 8 \\
SN~1995V & IIP     &  ~84~d    &       & 9 \\
SN~2001B  & Ib     &  ~60~d    &       & 7 \\
\hline
\end{tabular}
\end{center}
{\bf References:} (1) Meikle et al.~(1989);
(2) Spyromilio \& Leibundgut (1996);
(3) Fassia et al.~(2001);
(4) Spryomilio, Leibundgut \& Glimozzi (2001);
(5) Gerardy \& Fesen 2002, in prep.;
(6) Gerardy et al.~(2002, this paper);
(7) Gerardy et al. 2002, in prep.
(8) Wheeler et al.~(1994);
(9) Fassia et al.~(1998);
\end{table*}

\citet{spyromilio01} have concluded that CO formation is common in 
Type~II supernovae.  With the discovery of CO in the Type~Ic event SN~2000ew, we
have the first evidence that this conclusion might be extended to include 
core-collapse supernovae in general.  Table~\ref{sn00ew_tab1} lists all the 
core-collapse 
supernovae for which \textit{K}-band spectra have been published to date as well
as a couple of objects for which papers are in preparation.  For each object, we
list the epoch of the last non-detection of CO (if there were any), as well as 
the epoch that CO was first detected.  The epochs quoted are relative to the 
supernova discovery date or the date of the earliest pre-discovery detection.

Of the 10 objects with \textit{K}-band spectra, seven have CO detections.
Furthermore, it may be that the three non-detections were simply not observed
at a late enough epoch.  The results in Table~\ref{sn00ew_tab1} suggest that CO 
tends to
form around 100--200 days, and none of the three objects without detections were
observed past 100 days. 

\subsection{CO temperature}
A quantitative analysis of CO emission in SNe requires a good
understanding of the chemical and kinematic structure of the ejecta, as well as 
detailed modeling of the CO formation, destruction, and excitation processes.
(e.g.~\cite{gearhart99}.)  Unfortunately, such an analysis for SN~2000ew 
requires more information than is available in our single spectrum. Furthermore,
the S/N of the data does not place strong constraints on the fine-structure in 
the CO bands.  We will therefore restrict ourselves to a very basic analysis
of the CO emission along with a comparison to the spectrum observed in SN~1987A.

The shape of the CO profile provides a lower limit on the temperature of the
emitting region.  LTE modeling of CO has shown that for temperatures below
about 2000~K, the first two bandheads have comparable strengths, but higher
order bandheads are down by at least 20--30\% \citep{SH89,liu92}.  This
creates a noticeable break in the spectrum redward of the second bandhead 
near 2.32 \micron.  Since no break is evident in the spectrum of SN~2000ew,
we conclude that the temperature is at least 2000~K in the CO emitting region. 

\subsection{CO velocity}

\begin{figure*}[t]
\begin{center}
\includegraphics[scale=0.5,clip=true,trim=20bp 00bp 00bp 0bp,angle=270]{%
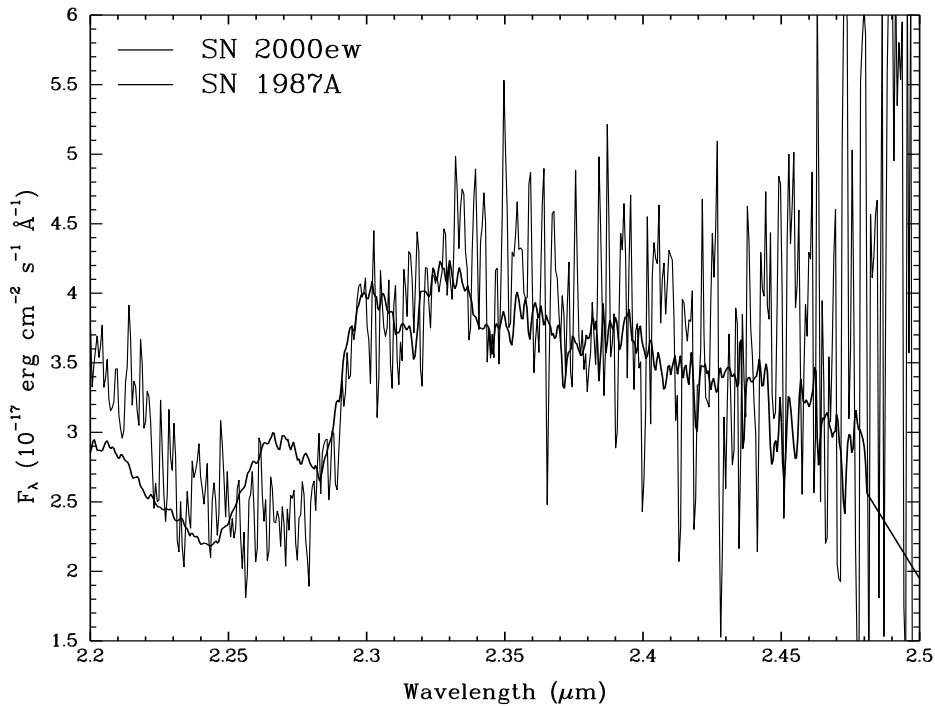}
\end{center}
\caption[Comparison of the CO detections in SN~2000ew and SN~1987A]{%
Comparison of the CO detections in SN~2000ew and SN~1987A.  The
SN~1987A has been shifted vertically and scaled to roughly match the continuum
and CO brightness of SN~2000ew.  The wavelengths of both spectra have been
shifted to the rest frames of the respective host galaxies.  The CO profile of
SN~2000ew appears quite similar to that of SN~1987A longward of 2.28~\micron,
but shows no trace of the unidentified emission feature seen in SN~1987A near
2.26~\micron.\label{sn00ew_fig2}}
\end{figure*}

It is useful to compare the observed CO emission in SN~2000ew with the much
higher S/N CO detection in SN~1987A.  Figure~\ref{sn00ew_fig2} shows the CO 
emission from 
SN~1987A on day 192 \citep{meikle89} overplotted on our spectrum of SN~2000ew.  
The SN~1987A data have been shifted vertically and scaled to match the continuum
and CO brightness of SN~2000ew.  Both spectra are plotted in the rest frame of 
the host galaxy.  

Longward of 2.28~\micron, the SN~2000ew data matches the SN~1987A data fairly
well.  The bandhead structure is not visible in SN~2000ew but this is likely
due to the low S/N of the SN~2000ew detection rather than velocity smearing.  
In fact, the blue edge of the CO feature in SN~2000ew matches that in SN~1987A 
quite well, suggesting that the CO has a similar velocity in both objects: 
$\approx 2000$ km~s$^{-1}$ (e.g.,~\cite{liu92}).  

While a 2000 km~s$^{-1}$ CO velocity is not unexpected for a Type~II supernova
like SN~1987A, it is quite low for some hydrodynamic models of Type~Ic SNe.  
Spherically symmetric hydrodynamic explosion models of the Type~Ic SN~1994I
predict a minimum velocity for the C/O rich layer of around 7000 km~s$^{-1}$ 
\citep{iwamoto94}.  For such a model to be consistent with an observed CO 
velocity of 2000 km~s$^{-1}$ would require a kinetic energy release well below
$10^{51}$ erg.

On the other hand, the low CO velocity may be consistent with strongly 
non-spherical jet-driven explosion models.  In the asymmetric explosion models
used by \citet{maeda02} to analyse the Type~Ic hypernova SN~1998bw, ejecta 
velocities are large along the jet, but much more modest 
perpendicular to the jet axis. In their $E=3 \times 10^{52}$ erg model, the
C/O rich gas has a minimum velocity around 3000 km~s$^{-1}$.  With a lower
kinetic energy release such a model might be able to account for the low CO 
velocities observed in SN~2000ew. 

\subsection{An unidentified 2.26~\micron\ feature in SN~1987A} 
As shown in Figure~\ref{sn00ew_fig2}, SN~1987A exhibited an emission feature 
peaked near 
2.265~\micron\ which is absent in the SN~2000ew spectrum. To date, this line has
not been satisfactorily identified.  \citet{spyromilio88} identify the feature 
as CO$^+$ but detailed models of CO chemistry in SN~1987A have ruled out this 
identification \citep{liu95,gearhart99}.  Furthermore, this feature did not 
evolve with the CO bands in SN~1987A, remaining visible long after the CO 
emission had faded \citep{meikle93}. 

\citeauthor{spyromilio01} (\citeyear{spyromilio01}, hereafter SLG) suggest that 
the SN~1987A feature is ``ubiquitous'' in Type~II supernovae and 
must be accounted for when measuring the velocity of CO emission.  However, the
presence of this 2.26~\micron\ feature in other supernovae is somewhat 
uncertain.  Detections have been reported for SN~1995ad \citep{spyromilio96}, 
SN~1998dl, and SN~1999em (SLG).  However, in all 
three cases the purported detections are not clear features, but rather apparent
excesses of emission above the continuum level which depend critically on where 
the continuum baseline is defined in these low S/N spectra.  Furthermore, this 
feature is not seen in the CO spectrum of the Type~IIP SN~1999gi (Gerardy \& 
Fesen 2002, in prep.) nor is it seen in our CO spectrum of SN~2000ew, which 
otherwise shows many similarities to SN~1987A and other Type~II SNe.

\begin{figure*}[t]
\begin{center}
\includegraphics[clip=true,trim=0bp 0bp 40bp 0bp,scale=0.5,angle=270]{%
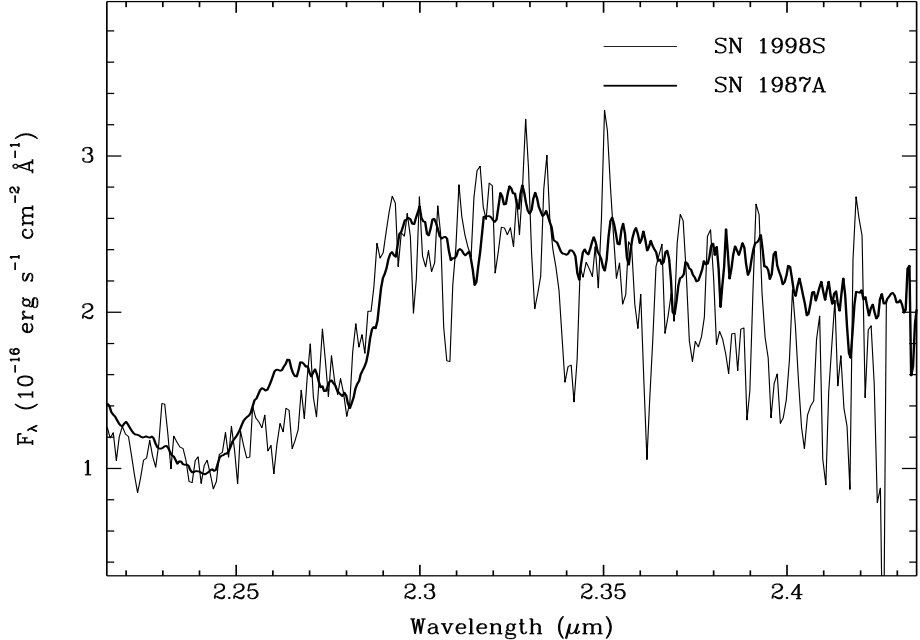}
\end{center}
\caption{Reproduction of Figure~6 of \citet{gerardy00} comparing the
CO emission in SN~1998S to that in SN~1987A.  The SN~1987A data has been shifted
vertically and scaled to roughly match the continuum and CO brightness of
SN~1998S.  The wavelengths of both spectra have been shifted to the rest frames
of the respective host galaxies.  The excess emission to the blue of the CO
bands in SN~1998S is not a good match to the unidentified emission
feature seen near 2.26~\micron\ in SN~1987A. The mismatch near 2.4~\micron\
may be due to residual telluric absorption, but could also be an indication of
a lower CO temperature in SN~1998S.\label{fig3}}
\end{figure*}

As evidence of the presence of the 2.26~\micron\ SN~1987A feature in other SNe, 
SLG reported that this feature was blended with the blue edge of the CO feature 
in the 20 June 1998 spectrum of SN~1998S presented by \citet{gerardy00} 
(hereafter GFHW).  SLG referenced Figure~6 of GFHW which is reproduced here as 
Figure~\ref{fig3}.  This figure shows the CO detection of SN~1998S with 
that of SN~1987A 
overplotted.  The SN~1987A spectrum has been shifted vertically and scaled to 
approximately match the continuum level and CO brightness of the SN~1998S 
spectrum, and the wavelengths of both spectra have been shifted to the rest 
frame of the supernovae.   
Near 2.4~\micron\, the SN~1998S spectrum is somewhat
fainter than the SN~1987A.  GFHW express some concern that this may be due
to residual telluric absorption in the data.  If real, the steeper profile is 
indicative of a somewhat lower temperature in SN~1998S compared to SN~1987A.

Both spectra show excess emission to the blue of the CO rise near 2.8~\micron.  
GFHW interpret the emission as the extended blue wing of a high velocity CO 
profile, but admit that there may be an unidentified line contributing to this 
emission.  They noted that the emission cannot be due to the 
2.26~\micron\ feature as that feature appears farther to the blue in SN~1987A.  
SLG reached the opposite conclusion, stating that the emission in SN~1998S was 
the same feature as that seen in SN~1987A, and attributed the difference in 
wavelength to the different recession velocities of the supernovae.  However, 
since both spectra in Figure~\ref{fig3} (and Figure~6 of GFHW) have been 
Doppler shifted 
to the supernova rest frame, we do not agree with their conclusion. 

A SN~1998S spectrum obtained by \citet{fassia01} casts further doubt on the 
presence of this 2.26~\micron\ feature.  In their spectrum taken one day before 
that of GFHW they find little excess emission blueward of the sharp edge of the 
CO bands and suggest that the differences between the two spectra are largely 
due to noise. 

Consequently, we suggest that the SN~1987A feature is not ubiquitous, and may 
in fact be relatively rare, with SN~1987A remaining the only unambiguous 
detection of this feature.  

\section{Conclusions}
We have presented \textit{K}-band spectra of the Type~Ic SN~2000ew which shows
strong first-overtone emission of carbon monoxide.  This is the sixth 
core-collapse supernova for which a CO detection has been published and it is 
the first detection in a non-Type~II supernova. 

The spectrum also shows narrow emission lines of [\ion{Fe}{2}] and \ion{He}{1} 
but is conspicuously lacking narrow \ion{H}{1} (Br$\gamma$) emission.  
We interpret
this emission as coming from dense clumps of hydrogen-poor circumstellar gas.

\citet{spyromilio01} have concluded that CO formation probably occurs in all 
Type~II supernovae.  We present a table listing ten core-collapse SNe for which 
\textit{K}-band spectra were obtained, seven of which show CO emission.  CO 
emission was typically first observed between 100 and 200~d, and the three SNe 
without CO detections were not observed past 100~d.  

The detection of CO in the Type~Ic SN~2000ew provides the first indication that 
CO formation may also be common in stripped-envelope core-collapse SNe (Types 
Ib, Ic, and ``IIb'').  The observed CO emission profile in SN~2000ew is quite
similar to that of SN~1987A, and indicates an expansion velocity around 
2000 km~s$^{-1}$, which is surprisingly low for a Type~Ic supernova.  The low
observed CO velocity may be another indication that strongly non-spherical
explosion models are needed for Type~Ic SNe. 

To date, there have been no observations that provide strong evidence for high
velocity CO emission.  An apparent blue wing in the CO spectrum of SN~1998S 
presented by GFHW might be due to high velocity CO. However a coeval spectrum 
taken by \citet{fassia01} doesn't show this blue extension to the CO profile and
they derive a much lower CO velocity from their observations.  Thus the CO
velocity in SN~1998S is uncertain. In all of the other CO detections, the 
observed velocity is around 2000 km~s$^{-1}$ or lower.  

It could be that a low C/O velocity is a requirement for CO formation.  
CO formation is a highly density sensitive process, and it is possible that in 
higher velocity gas the density decreases too fast to allow much CO to form.  
However, for most type~II supernovae, the C/O rich layers are buried too deep 
in the ejecta to expect CO velocities much
higher than 2000 km~s$^{-1}$.  Looking for CO emission in type~Ib/c supernovae 
will provide a better test as much higher C/O velocities are expected in these
objects. 

We find no evidence in SN~2000ew for the presence of a 2.26~\micron\ feature 
seen in SN~1987A and, contrary to \citet{spyromilio01}, we suspect it is either 
faint or absent in other Type~II and Type~Ib/c objects.  SN~1987A remains 
the only clear detection of this feature, and its identification remains a
mystery.

~\\
We would like to thank the Subaru observatory staff for their excellent support,especially Dr.~Hiroshi Terada.  C.~L.~G. and R.~A.~F.'s research is supported
by NSF grant 98-76703.  K.~N. has been supported in part by the Grant-in-Aid forScientific Research (07CE2002, 14047206, 14540223) of the Ministry of Education,Culture, Sports, and Technology in Japan.  Research of JCW is supported by NSF
Grant 0098644.
~\\

\end{document}